# Mobile Social Big Data: WeChat Moments Dataset, Network Applications, and Opportunities

Yuanxing Zhang, Zhuqi Li, Chengliang Gao, Kaigui Bian, Lingyang Song, Shaoling Dong, Xiaoming Li

## Abstract

In parallel with the increase of various mobile technologies, the MSN service has brought us into an era of mobile social big data, where people are creating new social data every second and everywhere. It is of vital importance for businesses, governments, and institutions to understand how peoples' behaviors in the online cyberspace can affect the underlying computer network, or their offline behaviors at large. To study this problem, we collect a dataset from WeChat Moments, called WeChatNet, which involves 25,133,330 WeChat users with 246,369,415 records of link reposting on their pages. We revisit three network applications based on the data analytics over WeChatNet, i.e., the information dissemination in mobile cellular networks, the network traffic prediction in backbone networks, and the mobile population distribution projection. We also discuss the potential research opportunities for developing new applications using the released dataset.

## Introduction

In the past decade, the widespread use of social network has greatly enriched the daily lives of people, providing new forms of entertainment and constructing new types of relationships. According to a recent report from GWI on the daily time spent on social networks, people tend to spend more than two hours on social networking. Researchers from Stanford have proposed a network analysis and graph mining library, named SNAP (http://snap.stanford.edu/), to help computationally analyze the network structure.

Owing to the furious rate at which proprietary mobile technologies and networks evolve, the mobile social network (MSN) service arises with a new era of mobile social big data, where users could conveniently converse and connect with others, and create new social data every second, everywhere through their mobile devices [1]. Compared with the web-based social network, MSNs provide more opportunities for dedicated use in mobile and wireless networks, for example, location-based services, mobile communication, and augmented reality. However, due to the lack of promising datasets, only a few research works have focused on the new MSN.

## A Taxonomy of Mobile Social Network Applications

The interpersonal tie in social networks, a.k.a. the social tie, is defined as the information-carrying connection among social circles [2], with which they exchange and forward various kinds of information [3]. The strength of the social-tie between two people can be evaluated by how frequently they communicate with each other via any online and/or offline channel.

There are two broad categories of mobile social network applications, based on the strength of the social-tie between users.

**One-Way Following on MSN Among Strangers:** On Twitter or Weibo, following someone is not mutual. One can follow you without your approval (by default), and you do not have to follow them back. For example, a celebrity may have millions of Twitter or Weibo followers, most of whom they do not know in real life.

**Mutual-Following on MSN Between Friends:** On Facebook, two friends mutually follow each other if one approves the other's friend request. Meanwhile, Facebook also allows users to have one-way following of a page set up by businesses, organizations, and brands that allow everybody to follow. Meanwhile, social instant messenger applications, such as WhatsApp, WeChat,[1] and Line, require mutual-following between two acquaintances, and allow them to send instant messages to each other, instead of using the conventional SMS.

Two users with a mutual-following relationship usually have a stronger social tie than two users with only one-way following between them; two users connected by a social instant messenger may communicate with each other more frequently than those who only follow each other on the MSN without messaging. Many MSN applications have incubated their own messenger app, such as Facebook Messenger and direct message on Twitter.

## New Features of WeChat

In WeChat, friends not only send instant messages to each other, but also have access to each other's page, i.e., the page of "WeChat Moments" (WM), or the friend circle page. WM not only takes advantage of web-based social network services, but also adds the following new features that may delight users.

**Keep Strong Social Tie — No Access to Strangers' Page on WM:** The MSN and messenger apps

[1] WeChat is a popular mobile messenger app, and the "WeChat Moments" (a.k.a. the friend circle) provides WeChat users with the social-networking services like viewing and posting texts/photos/links on their pages.

Yuanxing Zhang, Zhuqi Li, Chengliang Gao, Kaigui Bian, Lingyang Song, and Xiaoming Li are with Peking University; Shaoling Dong is with Fibonacci Data Consulting Services Inc have defined different access policies to one's page, which affects the information diffusion process in various ways. On Twitter, one's page is open for anyone's access (by default). The path of retweeting is visible to everyone, and it is easy to follow anyone over the retweeting path. In WM, two users cannot see each other's page if they are not connected as friends, and there is no visible path of retweeting (reposting) a message, and no one has an idea about where the message is from.

**Selected Content Display — Private Content Displayed to Selected Friends:** In certain scenarios over WM, there exists content that would be better to be exposed only to selected close friends, which should not be viewed and reposted by other friends (e.g., no access to strangers' page). This kind of private content is only shared among selected close friends, which further strengthens the social tie among these friends.

**Group Chat — A Way of Approaching The Unfamiliars:** As an instant messenger, WeChat provides the group chat function for strangers to converse, as every user can join a group by clicking a group link, scanning a QR code, or being invited by an existing group member. Also, two members in the same group do not have to be friends (e.g., they are invited by two different members in this group). Users could send texts, pictures, video/voice clips, and links to HTML5 pages into the group, which has become a popular way to propagandize online/offline activities.

## Challenges in Data Analytics for Network Applications

Businesses, governments, and institutions are interested in data analytics to understand how peoples' online behaviors in the MSN can affect the underlying computer network, or their offline behaviors at large. However, new challenges may arise when we revisit the following problems, because the access control policy (e.g., no access to strangers' page, private content) in WM confines the information diffusion among "acquaintances," which reduces the chance that a post is exposed to strangers as well as the probability of reposting.

**Information Dissemination in Mobile Cellular Networks:** Information dissemination from one user to others in mobile cellular networks depends on the reliability of connections among cellular users (e.g., device-to-device connections). It is a challenging problem as we have no idea about the quality of connections among them. Meanwhile, we observe that information diffusion over the MSN receives a high probability of success, as MSN services can easily stimulate users to share information and messages with their loved ones while on-the-go. Hence, it will be interesting to investigate whether there exist influential users that have high-quality connections to help disseminate information in cellular networks.

**Backbone Network Traffic Prediction:** Offline locations of MSN users can be predicted by mining their periodic behaviors. Human movement and mobility patterns have a high degree of freedom and variation, but they can still exhibit structural patterns due to geographical and social constraints. Accordingly, user migration may cause a change in the backbone network traffic distribution, e.g., locations that have a high density of mobile users should be allocated with more backbone network resources. The traffic distribution of the backbone network depends on users' movement, and also correlates to how frequently they repost from friends. Online interaction frequency is an important indicator of the social tie among users in WM, which requires a fine-grained analysis to better allocate resources in the underlying backbone network.

**Projecting Mobile Population Distribution:** The online behaviors of users (e.g., reposting a link) may reflect certain geographical attributes. We could then conjecture that there lies a geo-homophily between an MSN and an offline mobile network. When large-scale migration (during Spring Festival in China, or winter/summer vacations) happens, changes in the structure of the social network could be monitored. Therefore, it is important for businesses to predict how the mobile population distribution varies, in order to deploy appropriate regional marketing strategies, or provide personalized recommendations for users at home, at work, or on travel.

In this article, we collect a dataset from WeChat Moments, called WeChatNet [4], which involves 25,133,330 WeChat users with 246,369,415 records of link reposting on their pages, from January 14 to February 27, 2016. This is the first released big dataset of WeChat Moments on users' reposting behaviors. We revisit three network applications based on the data analytics over WeChatNet. We first present a voting strategy that finds the most influential users for information dissemination in mobile cellular networks. By observing the interaction between friends in time and spatial domains, we predict the traffic load in the underlying backbone network with a prediction accuracy rate of over 90 percent, which yields a near optimal resource allocation (i.e., the server placement in the underlying backbone network). Based on the location of users who view and repost a link in WM, we propose a model to project the distribution of the floating population. We further discuss the potential research opportunities for developing new applications using the WeChatNet dataset we release.

## Dataset Collection

### WeChatNet Dataset

As mobile social networks are usually developed on mobile devices, conventional web pages may not provide good visual experience. HMTL5 is a promising way to adapt to the different screen sizes of mobile devices. WeChat provides several interfaces officially to help developers design HTML5 pages. Users can then easily get access to the page content, and repost those interesting pages to their friends through the MSN like WM.

In the WM network, the links shared/posted by users usually lead to a post in HTML5 (H5). Such WM posts provide users with interactive operations such as an online greeting card, a lightweight online game (e.g., flappy birds), psychological tests, and so on. A WM post can be released by the WM service provider (Tencent), or a third-party web developer.

| Service category | Service name | Twitter | Weibo | Facebook | WeChat | WhatsApp |
|---|---|---|---|---|---|---|
| Social networking services | Video sharing | ✓ | ✓ | ✓ | ✓ | × |
| | Personal page | ✓ | ✓ | ✓ | ✓ | × |
| | Post search | ✓ | ✓ | ✓ | ✓ | × |
| | Favorite post save | × | × | ✓ | ✓ | × |
| | **Access to pages of non-friends** | ✓ | ✓ | ✓ | × | n/a |
| Messenger services | Limit on # of followers/friends | No limit | No limit | 5000 | 5000 | No limit |
| | Video/audio chat | × | × | ✓ | ✓ | ✓ |
| | Group messaging | × | ✓ | ✓ | ✓ | ✓ |
| | Sending location | × | × | ✓ | ✓ | ✓ |
| | Voice messaging | × | × | ✓ | ✓ | ✓ |
| Miscellaneous services | Mobile payment | × | × | ✓ | ✓ | × |
| | Video games | × | ✓ | ✓ | ✓ | × |
| | **Offline services (taxi, ticket, etc.)** | × | × | × | ✓ | × |
| | Shopping | × | ✓ | ✓ | ✓ | × |

TABLE 1. Comparison of social network services: Twitter, Weibo, Facebook, WeChat, and WhatsApp. Statistics in the table are collected as of July 31, 2017.

We can see that Twitter and Weibo serve mostly as social networks, while WhatsApp serves mostly as an instant messenger; Facebook (Messenger) and WeChat have combined social networking and instant messengers. WeChat Moments appears as a mobile social network with many features.

**WM Data Collection:** Our goal is to collect the statistics of WM post diffusions. We use the Application Programming Interface (API) provided by a business WM page creator platform, FIBODATA (http://www.fibodata.com/), for crawling diffusion trails of pages created over the platform. Based on the collected data, we are able to construct a diffusion graph for each posted WM page. The dataset contains about 320,000 pages created by businesses from January 14, 2016 to February 27, 2016, which involves 25,133,330 Wechat users with 246,369,415 link retweeting records.[2]

**Information Diffusion Process:** Suppose that users *i* and *j* are friends in the WM network. When user *i* shares the link of a WM post with his friends, user *j* may click the shared link to view the content of this post. If user *j* finds this post interesting, he may further repost the link to his friends, so that more users would have the chance to view this WM post. As this process is similar to the spread of infection:

- We call a user an *infected* user of a WM post if he views the post.
- We call a user an *infectious* user of a WM post if he views the post and reposts the post link.

**Post View Record:** A post view record in our dataset is a 5-tuple in the following format:

$$<U_1, U_2, PID, IP, t>,$$

where $U_1$ is the ID of the user whose post gets viewed; $U_2$ is the ID of the user who views the post by user $U_1$; PID is the ID of the WM post that is assigned by the WM post creator platform; *IP* is the IP address of the post viewer, that is, user $U_2$'s IP address; and *t* is the time when the post view happens. The whole tuple records the post view event that user $U_2$ at the address *IP* at time *t* views post *PID* of user $U_1$.

## Comparison of MSN Applications

In Table 1, we compare five most representative social network applications, including instant messengers. We can clearly see that Twitter and Weibo serve mostly as social networks, while WhatsApp serves mostly as an instant messenger; Facebook (Messenger) and WeChat have combined both social networking and instant messengers. WeChat Moments appears as a mobile social network with many noteworthy features.

First, WM takes advantage of the influential users (accounts) in the social network, and proposes *Subscription Accounts*, where companies or individuals could edit web pages and push to the followers. This function has been proved to have great impact to attract users' attention. Second, WM inherits the benefits of strong social ties of the instant messenger. The relationship between users in instant messenger (WeChat) is bidirectional rather than one-way following (Weibo). The relationship is also private, which means that strangers have no access to the information of certain users if they are not friends (i.e., no access to stranger's pages). This could be a double-edged sword for commercials, for instance, the advertisement may profit from the communication over strongly-tied friends, while the advertisement may have the difficulty of being spread widely and the business cannot obtain a long reposting path in most cases. Third, WeChat exploits the superiority of group chat. The HTML5 page with abstract content instead of the URL address can be posted into the group chatting window, making it easier for users to acquire information. Fourth, WeChat is also unique in its connection to offline services, like ordering food delivery, calling a cab, obtaining coupons by scanning QR-code, and so on.

[2] We release a WM dataset over https://github.com/pkumobile/WMdata. This dataset will be updated by crawling more recent data via FIBODATA.

In this article, we study the impact of online reposting behaviors of users over the underlying computer networks, and we showcase three typical network applications using the WeChatNet dataset, as illustrated in Fig. 1.

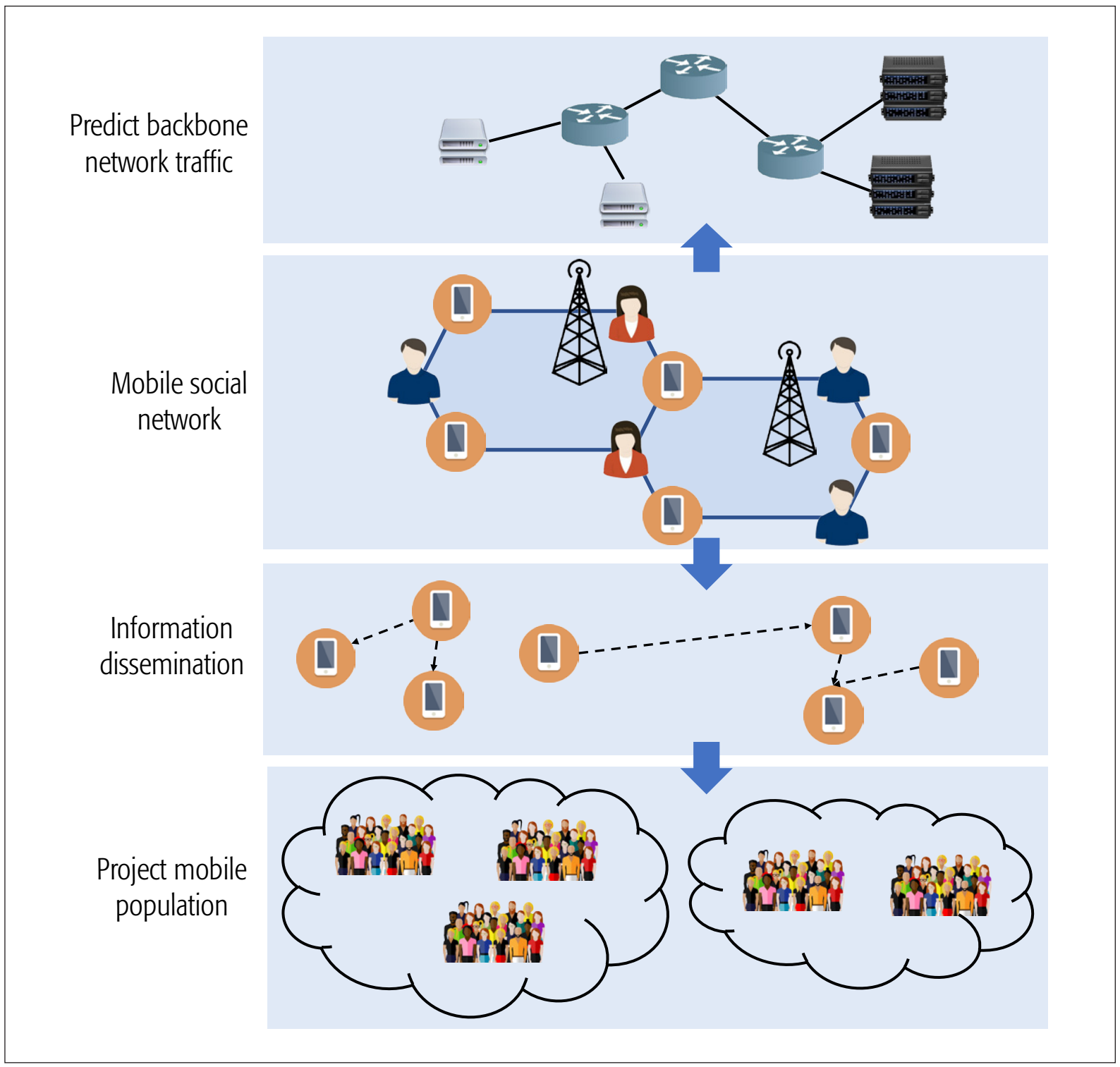

FIGURE 1. Three typical network applications based on the data analytics over the WeChatNet dataset.

## Information Dissemination in Mobile Cellular Networks

The reposting behaviors of users can form an information diffusion graph, and we could analyze the social influence over the network connections among users, which helps information dissemination in mobile cellular networks.

### Selecting the Most Influential User

The quality of experience in mobile cellular networks can be largely affected by the reliability of connections among cellular users/devices (e.g., device-to-device connections) [5, 6]. For example, the more high-quality connections one has, the more-widely they can spread the information to other users/devices. In this study, we are interested in finding the most influential user who can best help disseminate information in mobile cellular networks.

In the same geographical community, the connection graph of mobile cellular users can be built on the basis of the information diffusion graph in the mobile social network. Hence, we can transform the problem of selecting the most influential user in mobile cellular networks to that of finding key opinion leaders (KOLs) in mobile social networks.

### Implementation by the Voting Strategy

The KOL with millions of followers (or a large number of friends) shows a strong social tie in the MSN, and they may have the power to help spread information for online marketing/advertising. For example, businesses usually invite selected consumers to advise the product by trials or rewards through Subscription Accounts of WeChat. These consumers should be KOLs so that the influence of the advertisement could be maximized. A user with millions of followers on Twitter or Weibo should be a KOL. However, in WM, it is challenging to detect a KOL by counting the number of their friends, because most users in WM only have a limited number of friends, and WM has imposed a maximum number of friends that one can connect to. Even a celebrity may only have approximately hundreds of friends in WM.

The information diffusion process of a mobile social network can be abstracted by an Independent Cascade (IC) model [7]: when a user becomes active (they receive a message), they have a single chance of activating each currently-inactive neighbor (who has not received the message) with a probability determined by interests, relationships, and so on. Instead of counting the number of their friends, we present a voting-based strategy that selects influential nodes by looking at the user's local contribution to the information diffusion process, which is different from existing greedy solutions using Lazy Evaluation strategies that tend to seek nodes with a large marginal influence. The voting-based strategy works as follow. First, it tries to identify those major diffusion trees for the information diffusion process. Then, in each diffusion tree, each offspring node can vote for its father nodes as the WM page is diffused from the father nodes to the offspring node. After repeating these steps multiple times, the number of votes received by a node in the diffusion tree indicates the influence of the node in the network.

In WM, the size of the diffusion graph/tree could be very large, and we cannot enumerate all diffusion trees or nodes for executing the voting operations [7]. Hence, we leverage the Gibbs Sampling technique to complete the above voting process within a limited time period. There are two parameters in this strategy, simply speaking, we need to sample a number of $R_1$ diffusion trees; for each selected tree, we execute a number of $R_2$ voting operations. Let $S$ denote the set of $K$ influential users that we find by using the voting strategy, and let $\sigma(S)$ denote the calculated influence of the set $S$.

We use the data on January 14, 2016, and Fig. 2a shows the influence value of $S$ under the voting strategy by varying $R_1$, while $K = 100$, $R_2 = 100000$. The green line represents the performance of the conventional greedy algorithm that chooses the user who will bring the greatest marginal value (the greatest increase in $\sigma(S)$) as the KOL, when $K = 100$. We observe that as $R_1$ grows, the variance of $\sigma(S)$ starts decreasing until it reaches a stable value. The influence value of $S$ under the voting strategy is greater than that under the greedy one, when $R_1$ is greater than 200. Given $K$ = 100, $R_1$ = 500, the σ(*S*) rises as $R_2$ increases, while the variance implies that the enumeration is quite random if $R_2$ is small. Similar to results in Fig. 2a, the influence value of *S* reaches a stable value under the voting strategy when $\log_{10}(R_2)$ is greater than 4.

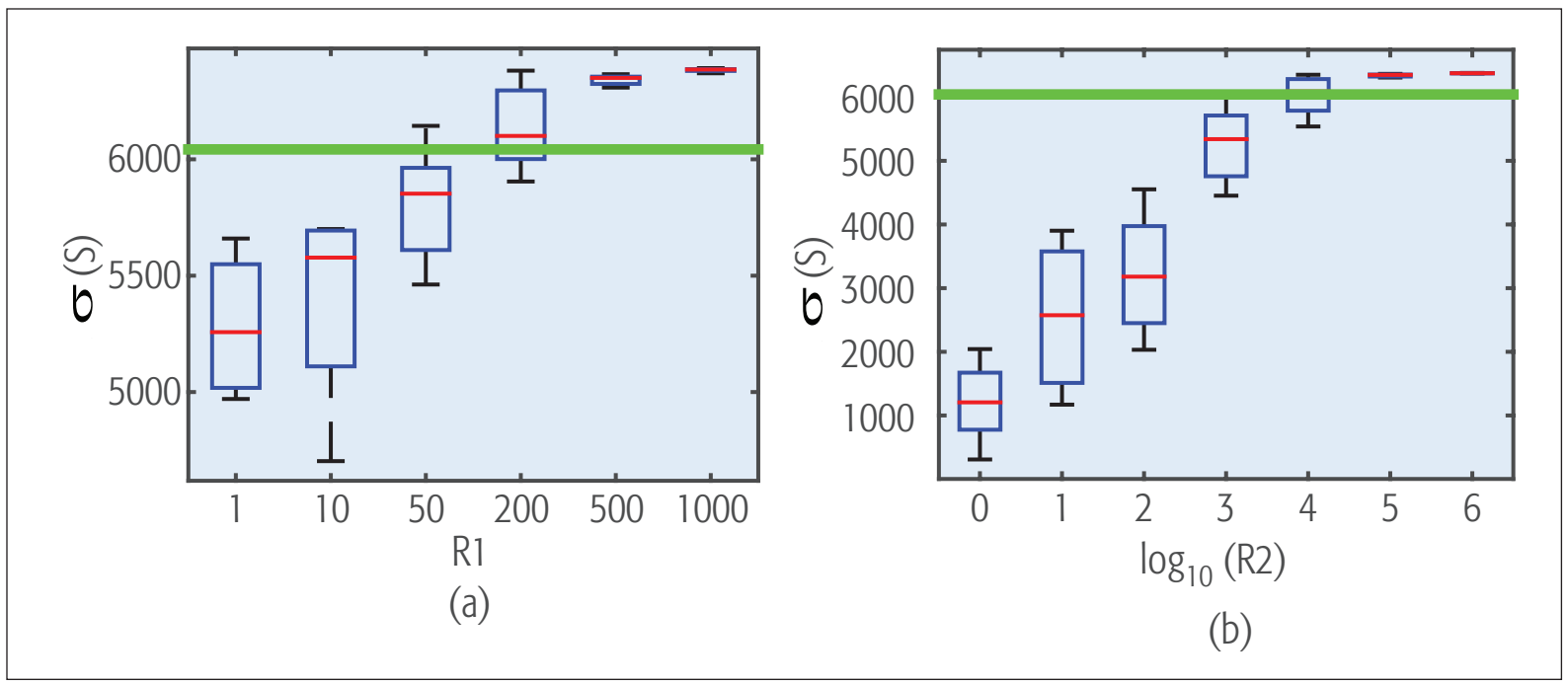

FIGURE 2. The experimental results of the voting-based strategy in finding the most influential users: a) σ(*S*) vs. $R_1$, the number of samples diffusion trees; b) σ(*S*) vs. $R_2$, the number of votes for each tree.

## Profiling Traffic Distribution in Backbone Networks

People prefer communicating with each other, sharing videos/images with friends, through MSN like WeChat. Therefore, mobile social network messaging could partially reflect network traffic. By observing the interaction between friends over WeChat in time and spatial domains, it is feasible to predict the traffic load in the underlying network, and present a reverse greedy strategy to better place servers and balance the traffic load in the network.

### From Communication Pattern to Traffic Prediction

We model the communication pattern of WeChat users as a Markov random field and compute the communication distance between two users. With communication distance and frequency, we can predict the traffic generated by an online MSN in the underlying network.

To validate the model, we utilize the data of the first 19 days in the dataset as the training set to train the model. The data in the following five days are used for testing the performance of the prediction model. Figures 3a and 3b show the real/predicted communication distance, and the real/predicted traffic load, where we observe that the model could do the prediction quite well, with an error rate lower than 10 percent.

### Traffic Optimization and Server Placement

We then design a heuristic approach, called the reverse greedy strategy, to ease the traffic load generated by WeChat communications. The goal of traffic optimization is to reallocate network resources (the servers) at proper locations to shorten the communication distance between users. Intuitively, we can select the location that can bring the most reduction in the generated traffic load in a greedy manner.

We use the backbone network graph of China Unicom and China Telecom to emulate the performance of the reverse greedy strategy. We enumerate all the possible combinations of server placement in the graph, calculate the resulting traffic load, and choose the case with the least load as the "optimal" solution. Figures 3c and 3d show that the reverse greedy strategy can achieve the near optimal result in the two backbone network graphs, while the performance of the naive greedy strategy is inferior. In addition, the curves for the optimal and the reverse greedy strategy become flat gradually after placing five servers, which implies that replacing five servers is sufficient to reduce most of the traffic. We also visually display the results of server replacement of the two greedy strategies and the optimal solution in Figs. 3e, 3f, and 3g, where the result of the reverse greedy strategy is quite similar to that of the optimal solution, and the naive greedy strategy fails to consider the nationwide replacement of servers.

## Projecting Mobile Population Distribution

Projecting the population distribution in geographic regions is important for many applications such as launching marketing campaigns or enhancing public safety in certain densely populated areas. Conventional studies require the collection of people's trajectory data through offline means, which is limited in terms of cost and data availability. The wide use of MSN apps over smartphones has provided the opportunity to devise a lightweight approach of conducting the study using the online data of smartphone apps [8].

### Modeling Geo-Homophily

A division of geographical regions is stable only if the MSN users in these divided regions show a strong geo-homophily, i.e., people in each region prefer communicating with others in the same region more than with those in other regions. These inspire us to investigate the relationship between online information diffusion, that is, users' communication in MSN, and the population distribution over a fixed division of offline regions. Intuitively, we can use the geo-location of messages among MSN users to derive the user distribution over the given regions. Then, the floating population across regions can be further inferred based on the derived distribution, which could be explained by the Dirichlet Process Mixture.

### Geo-Homophily in WM

The dataset records the users' re-tweeting in 34 provinces in China, and we use these provinces as the geographic regions in this experiment. Every user in WM should have viewed a collection of pages, and each page view's IP address resides in a province, among which the most frequently recorded one is set as the province where the user is located. We analyze the message diffusion process in two time periods. Before Spring Festival, we monitor the message diffusion from January 14 to January 31, 2016, which includes pre-holiday working and weekend days. On the Spring Festival day, most people stay at home, and hence the structure of the message diffusion graph would be different.

Figure 4a shows the volume of message diffusion inside each province and that between every pair of provinces of China, where the amount of message diffusion inside a province is proportional to the size of the corresponding circle, and the amount between provinces is represented by the length of arcs.

The results indicate that most of the diffusions occur inside provinces, so the arcs are relatively sparse. In particular, there lies no arc between some pairs of circles in this figure, which does not mean that there is no message diffusion between the corresponding two provinces, but implies that the message diffusion between them is much weaker than that between those pairs of circles having arcs. For example, there were only hundreds of message diffusions between Tibet and Taiwan in the dataset; in contrast, several millions of message diffusions occur between Beijing and Guangdong. This can be explained by the fact that those provinces are at distant locations, or they have little communication with most provinces in mainland China.

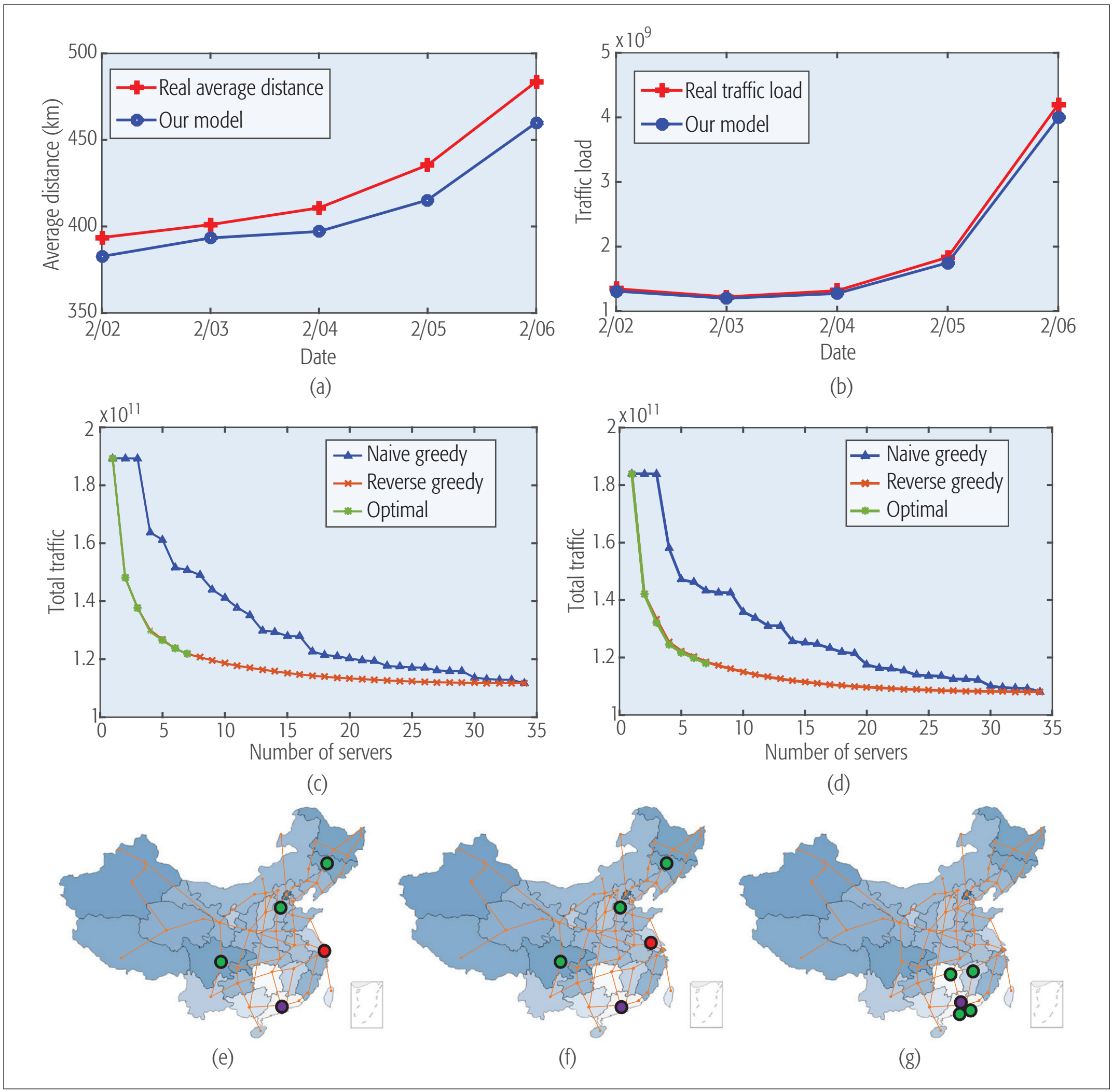

FIGURE 3. Evaluation results on network traffic optimization: a) the predicted average communication distance; b) the predicted traffic load; c) performance of the location selection strategy on China Unicom network; d) performance of the location selection strategy on China Telecom network; e) results of the optimal strategy for placing five servers; f) results of the reverse greedy strategy for placing five servers; g) result of the naive greedy algorithm for placing five servers.

On the Spring Festival holiday, most people stay with their families in their home province. The graph structure changes, as most of the messages are sent for appointments and greetings, and these diffusions mainly took place between friends in the same vicinity. Thus, the proportion of the diffusion inside regions increases. The graph structure is illustrated in Fig. 4b, where some inter-province arcs disappear.

Then we obtain a chaotic segregation, which can hardly be said to have any geo-homophily. The number of message diffusions inside and across these regions is shown in Fig. 4c. Compared to Fig. 4a, under the same scale of plotting, the distribution of circles representing regions is very dense, and most of circles' sizes are similarly small.

When considering the diffusion of a single page, we find that it will be reposted many times in the home region of the sender, while it may be sent to only a few non-home regions; those diffusions across regions take up only a small part. For example, in Fig. 5a we illustrate the distribution of views to a message with approximately one million views, where the message is originally sent from the region of Beijing.

We then evaluate the performance of the Dirichlet Process on the WeChatNet dataset, and compare it against the results of the latest national population census in China (http://www.stats.gov.cn/tjsj/pcsj/

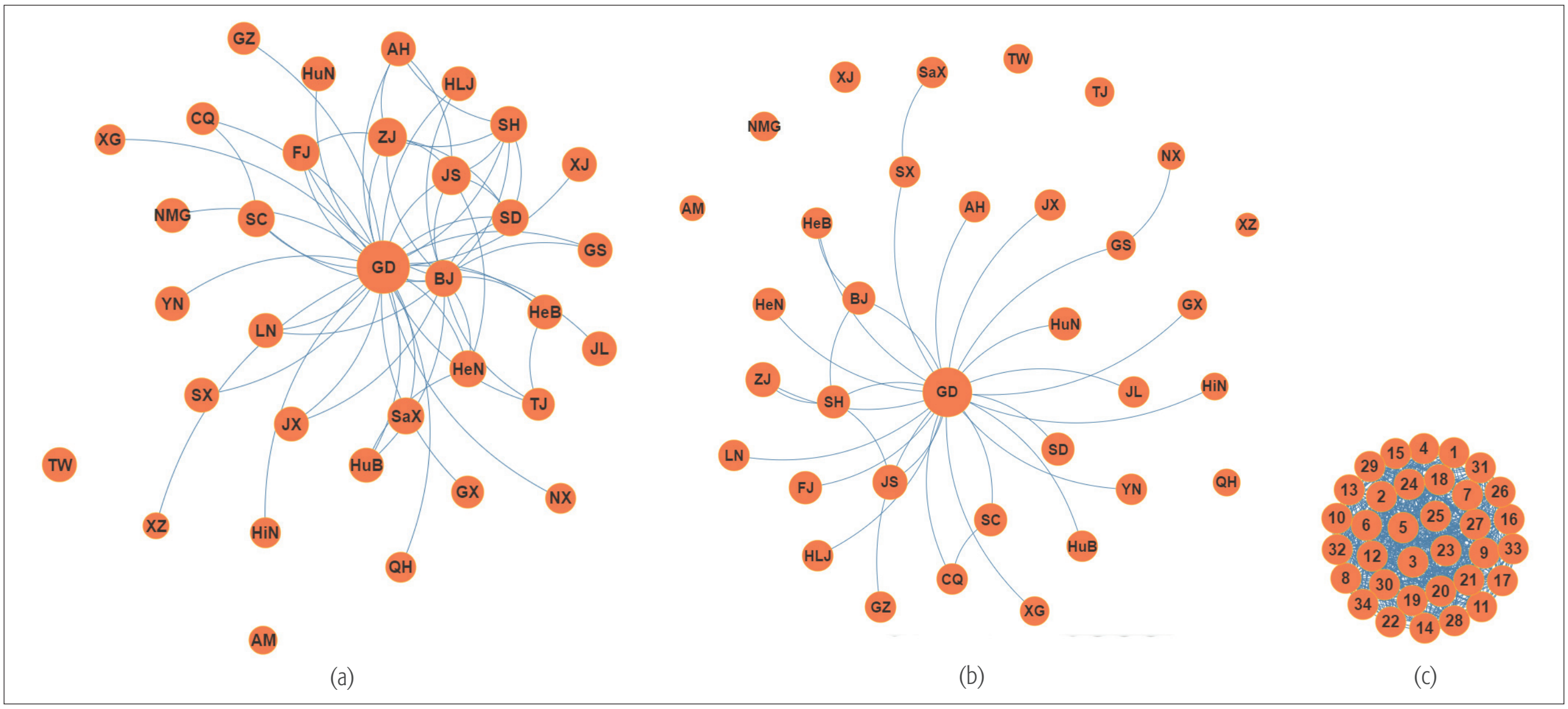

FIGURE 4. The visualization of the Geo-homophily on three diffusion graphs: a) before Spring Festival; b) on the Spring Festival day; c) baseline network.

rkpc/6rp/indexch.htm), which provides us the statistics of floating population in China. Here, the floating population (FP) in our experiments has excluded those whose home and remote regions are the same (e.g., those who rarely move out of the home region, as the work place and the home belong to the same region). As shown in Fig. 5b, there lies a linear correlation between the real floating population and the predicted value by the Dirichlet Process.

## Research Opportunities

**Structure of New Social Networks:** We have discussed that the structure of the new MSN (e.g., WM) is different from that of the blog-like social network. It would be interesting to investigate in more depth the reasons why WM becomes different from other social networks. The data of WM provides new opportunities to explore the evolution of this new type of social network. For instance, the network representation of WM can recognize the similarity between users and predict potential links [9] in the social network that may not be easily obtained in the diffusion graph.

**Marketing in Dynamic Diffusion Graph:** Marketing is the most prevalent business model in online social network services. Our previous works have analyzed the results in the static diffusion graph. However, in a longer time period, the role or the influence of users in the social network may change, and it would be interesting to study the social tie between friends in a dynamic diffusion graph. Moreover, the analysis on the effects of extrinsic rewards (e.g., the advertisement viewing duration), and on friend-sourcing [10] could contribute to the marketing strategy as well.

**Understanding More Demographic Results:** WM has been proven to have the ability to profile traffic distribution and to project population distribution, and indeed, the dataset can be used to understand more demographic results by observing the interaction among people, such as how to identify a user in the social network with incomplete knowledge about his friends [11].

**Spam Detection:** The MSN facilitates the ubiquitous access to news/messages for everybody. However, rumors and cyber-violence are also easy to spread, breaking the right of legal citizens. We could analyze the network elements, that is, the influence and vulnerability of users (by resembling mean opinion score) in the group and the confidence of links (by resembling stability), to help detect seditious news/messages [12], and prevent them from spreading.

**Privacy Leakage:** Along with the wide spread of mobile social networks, there is an increasing concern about privacy issues. The user interests, locations and other kinds of private information can be easily exposed through data analytics [13], and even the social circles of people are vulnerable [14]. The WeChatNet dataset can be used to determine which kinds of private information might be leaked over the current domains of data, and which kinds of strategies could be leveraged to degrade the leakage of private information.

**Promoting Offline Marketing Activities:** Businesses have seen WeChat as an important offline marketing tool, as HTML5 pages redirected by scanning QR-codes can provide more information to consumers. Businesses have to determine the qualification of treating certain potential consumers as a group by observing the stability and the activity of the group. This could help them decide whether to launch marketing in a certain area during a specific time period. Device-to-device and vehicular communication technologies are found to be efficient in sharing information among users [15], which can greatly extend information diffusion in WM via either online or offline channels.

## Conclusion

In this article, we present a systematic study to understand how online reposting behaviors of users in WeChat will affect computer networks and users' offline behaviors. From the perspective of information diffusion, we present an online KOL detection method that is independent of the number of one's friends. By observing the interaction between friends, we predict the traffic load in the underlying network with a prediction accuracy rate of over 90 percent, which leads to a near optimal resource allocation. We also find that the Dirichlet Process Mixture can describe the distribution of the floating population, based on which we propose a model using online message diffusion to project the offline population distribution. Moreover, we point out the future research opportunities of using the dataset we release in understanding the impact of online social ties over the offline social, business, and political lives.

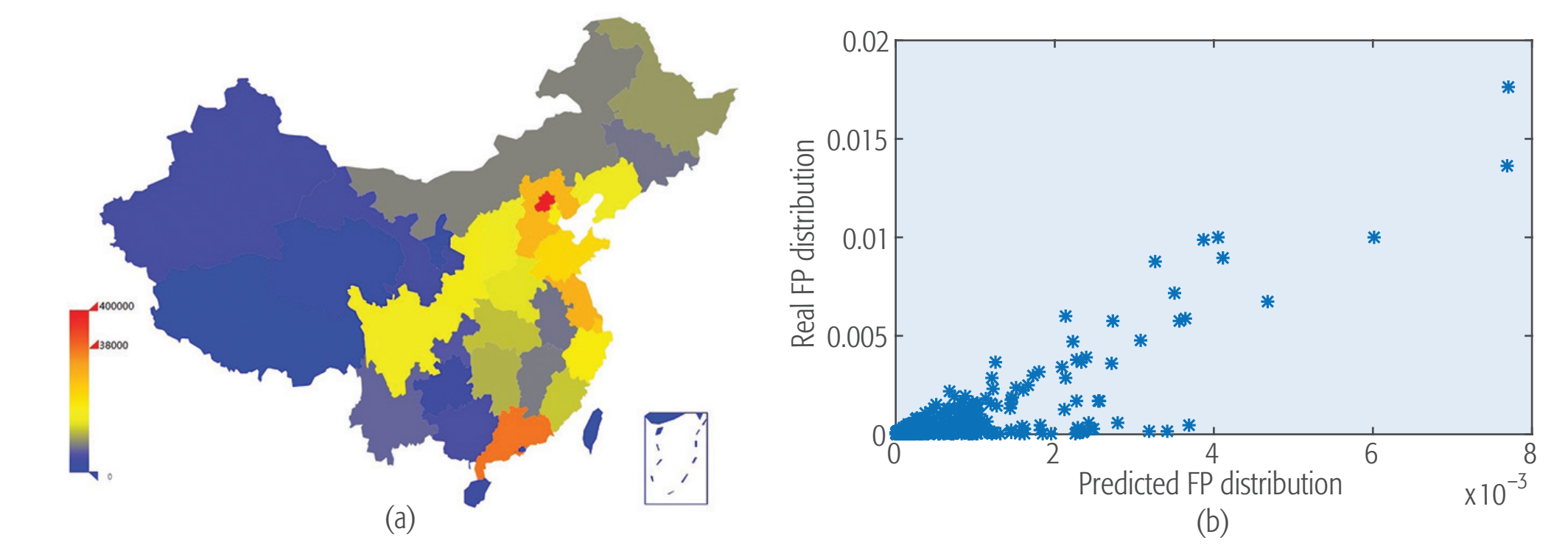

**FIGURE 5.** a) The viewing distribution of a message originated from Beijing; b) results of inferring the floating population (FP) that has excluded those whose home and remote regions are the same.

## Acknowledgment

This work is partially supported by the National Key Research and Development Program No. 2017YFB0803302, the National 973 Grant No. 2014CB340405, and the National Natural Science Foundation of China under Grant Nos. 61572051 and 61632017.

## References

[1] S. Ioannidis, A. Chaintreau, and L. Massoulie, "Optimal and Scalable Distribution of Content Updates over a Mobile Social Network," *Proc. IEEE INFOCOM 2009*, IEEE, 2009, pp. 1422–30.

[2] J.-P. Onnela *et al.*, "Structure and Tie Strengths in Mobile Communication Networks," *Proc. National Academy Sciences*, vol. 104, no. 18, 2007, pp. 7332–36.

[3] K. Wei *et al.*, "CAMF: Context-Aware Message Forwarding in Mobile Social Networks," *IEEE Trans. Parallel Distributed Systems*, vol. 26, no. 8, 2015, pp. 2178–87.

[4] WeChatNet, "A Mobile Social Big Dataset of WeChat Moments," https://github.com/pkumobile/WMdata.

[5] J. Liu *et al.*, "Device-to-Device Communications for Enhancing Quality of Experience in Software Defined Multi-Tier LTE-A Networks," *IEEE Network*, vol. 29, no. 4, 2015, pp. 46–52.

[6] J. Liu *et al.*, "Device-to-Device Communication in LTE-Advanced Networks: A Survey," *IEEE Commun. Surveys & Tutorials*, vol. 17, no. 4, 2015, pp. 1923–40.

[7] Y. Zhang *et al.*, "Influence Maximization in Messenger-Based Social Networks," *Proc. 2016 IEEE Global Commun. Conf. (GLOBECOM)*, IEEE, 2016, pp. 1–6.

[8] Y. Zhang *et al.*, "Population Distribution Projection by Modeling Geo Homophily in Online Social Networks," *Proc. 2nd Int'l. Conf. Crowd Science and Engineering (ICCSE'17)*, ACM, 2017, pp. 1–8.

[9] Z. Wang, C. Chen, and W. Li, "Predictive Network Representation Learning for Link Prediction," *Proc. 40th Int'l. ACM SIGIR Conf. Research Development Information Retrieval*, ACM, 2017, pp. 969–72.

[10] H. Zhu *et al.*, "A Market in Your Social Network: The Effects of Extrinsic Rewards on Friendsourcing and Relationships," *Proc. 2016 CHI Conf. Human Factors in Computing Systems*, ACM, 2016, pp. 598–609.

[11] M. Kearns *et al.*, "Private Algorithms for the Protected in Social Network Search," *Proc. National Academy of Sciences*, vol. 113, no. 4, 2016, pp. 913–18.

[12] J. Sampson *et al.*, "Leveraging the Implicit Structure within Social Media for Emergent Rumor Detection," *Proc. 25th ACM Int'l. Conf. Information Knowledge Management*, ACM, 2016, pp. 2377–82.

[13] M. Li *et al.*, "All Your Location are Belong to Us: Breaking Mobile Social Networks for Automated User Location Tracking," *Proc. 15th ACM Int'l. Symposium Mobile Ad Hoc Networking and Computing*, ACM, 2014, pp. 43–52.

By observing the interaction between friends, we predict the traffic load in the underlying network with a prediction accuracy rate of over 90 percent, which leads to a near optimal resource allocation.

[14] H. Zhu *et al.*, "Fairness-Aware and Privacy-Preserving Friend Matching Protocol in Mobile Social Networks," *IEEE Trans. Emerging Topics Computing*, vol. 1, no. 1, 2013, pp. 192–200.

[15] H. Nishiyama *et al.*, "Relay by Smart Device: Innovative Communications for Efficient Information Sharing Among Vehicles and Pedestrians," *IEEE Vehicular Technology Mag.*, vol. 10, no. 4, 2015, pp. 54–62.

## Biographies

Yuanxing Zhang [S'16] (longo@pku.edu.cn) is a Ph.D. student at the School of Electrical Engineering and Computer Science, Peking University, China. He received his B.S. degree in computer science from Beijing University of Technology in 2015. His current research interests include AI-assisted resource allocation and recommender systems.

Zhuqi Li (lizhuqi@pku.edu.cn) is a Ph.D. student in the Computer Science Department at Princeton University, USA. He received his B.S. degree in computer science from Peking University in 2017. His current research interests include social networks and mobile sensing.

Chengliang Gao [S'16] (gaochengliang@pku.edu.cn) is a master student at the School of Electronics Engineering and Computer Science, Peking University, China. His main research interests include mobile computing and social networks.

Kaigui Bian [S'05, M'11] (bkg@pku.edu.cn) is an associate professor at the School of Electronics Engineering and Computer Science, Peking University, China. His main research interests include mobile computing, cognitive radio networks, network security and privacy. He received the best paper award at IEEE ICC 2015 and ICCSE 2017. He is a recipient of 2014 CCF-Intel Young Faculty Researcher Award.

Lingyang Song [S'03, M'06, SM'12] (lingyang.song@pku.edu.cn) is a professor at the School of Electronics Engineering and Computer Science, Peking University, China. His main research interests include MIMO, cognitive and cooperative communications, physical layer security, and wireless ad hoc/sensor networks. He is the recipient of the 2012 IEEE Asia Pacific Young Researcher Award and the 2016 IEEE ComSoc Leonard G. Abraham Prize. He is currently on the editorial board of *IEEE Transactions on Wireless Communications*. He has been an IEEE Distinguished Lecturer since 2015.

Shaoling Dong (dong@szzbmy.com) holds a juris doctor degree from Peking University, China. He is the founder of Rabbitpre.com and Fibonacci Data Consulting Services Inc., Shenzhen, China. His research interests include communications and social network. He has been listed among the 2017 Forbes China 30 business leaders under 30 years old, and he was an invited speaker at Cannes Lions 2017.

Xiaoming Li [SM'05] (lxm@pku.edu.cn) graduated from Stevens Institute of Technology, New Jersey, with a Ph.D. in computer science. He is now a professor at Peking University. His research interests include search engines and web mining. He is a Fellow of the Computer Federation of China and a member of Eta Kappa Nu. He serves on the editorial board of *Concurrency and Computation* (Wiley). He received the CCF Wang Xuan Award and the Outstanding Educator Award in 2013 and 2014, respectively.